\documentclass[11pt]{amsart}

\usepackage{amsmath, amsthm, url, epsfig}
\usepackage{amssymb, amsfonts}
\usepackage{graphicx}
\usepackage{geometry}
\usepackage{soul}
\usepackage{xcolor}
\usepackage{subfig}
\usepackage[export]{adjustbox}
\usepackage{arydshln}
\usepackage{verbatim}

\usepackage{parskip}
\usepackage{fullpage}
\begin{document}

\title{Ranked Choice Voting And Condorcet Failure in the Alaska 2022 Special Election: How Might Other Voting Systems Compare?}
\author{Jeanne N. Clelland}
\address{Department of Mathematics, 395 UCB, University of
Colorado, Boulder, CO 80309-0395}
\email{Jeanne.Clelland@colorado.edu}

\thanks{The author was partially supported by a Collaboration Grant for Mathematicians from the Simons Foundation.}

\begin{abstract}

The August 2022 special election for the U.S. House of Representatives in Alaska featured three main candidates and was conducted by the single-winner ranked choice voting system known as ``Instant Runoff Voting." The results of this election displayed a well-known but relatively rare phenomenon known as  ``Condorcet failure:"  Nick Begich was eliminated in the first round despite being more broadly acceptable to the electorate than either of the other two candidates.  More specifically, Begich was the {\em Condorcet winner} of this election: Based on the Cast Vote Record, he would have defeated each of the other two candidates in head-to-head contests, but he was eliminated in the first round of ballot counting due to receiving the fewest first-place votes.

The purpose of this paper is to use the data in the Cast Vote Record to explore the range of likely outcomes if this election had been conducted under two alternative voting systems: Approval Voting and STAR (``Score Then Automatic Runoff") Voting.  We find that under the best assumptions available about voter behavior, it is likely---but not at all certain---that Peltola would still have won the election under Approval Voting, while Begich would almost certainly have won under STAR Voting.

\end{abstract}


\maketitle

\section{Introduction}\label{intro-sec}

On August 16, 2022, Alaska held a special election to fill the seat of deceased U.S. House Representative Don Young.  For the special general election, there were three candidates: Democrat Mary Peltola and Republicans Nick Begich and Sarah Palin.  The election was Alaska's first statewide election conducted by Instant Runoff Voting (IRV), commonly referred to as Ranked Choice Voting (RCV).\footnote{``Ranked Choice Voting" is an umbrella term, referring to a variety of voting and tabulation systems for both single-winner and multi-winner elections.  In this paper, we will use the more precise term ``Instant Runoff Voting," which refers to the specific single-winner system used for the Alaska election.}  For this election, voters were allowed to rank all three candidates, and ballots were counted as follows:\footnote{Official results obtained from 
https://www.elections.alaska.gov/results/22SSPG/RcvDetailedReport.pdf, accessed Nov. 4, 2022.}
\begin{itemize}
\item Round 1: Only first-place rankings were counted.  The results of this round are shown in Table \ref{table:IRV-round-1}.  At the end of this round, the candidate with the fewest first-place votes (Begich) was eliminated.  Ballots on which Begich was ranked first were ``transferred" to their second-ranked candidate, if any.  Any ballot with no second-choice candidate indicated was considered ``exhausted" and was not counted in the second round.  Of the 53,810 ballots on which Begich was ranked first, 11,290 were exhausted and the remainder were transferred as shown in Table \ref{table:IRV-transfer}.

\item Round 2: After Begich's first-place ballots were transferred to their second-choice candidates, the votes were counted again.  The results of this round are shown in Table \ref{table:IRV-round-2}.  Since Peltola received more than 50\% of the votes still active in Round 2, she was declared the winner.

\end{itemize}

\begin{table}[h!]
\begin{center}
\begin{tabular}{|l|c|c|}
\hline
 {\bf Candidate} & {\bf Votes} & {\bf Percentage}\\
\hline
Peltola & 75,799 & 40.19\% \\
\hline
Palin  & 58,973 & 31.27\%  \\
\hline
Begich  & 53,810 & 28.53\% \\
\hline
\end{tabular}
\end{center}
\caption{First Round Special Election Results}
\label{table:IRV-round-1}
\end{table}

\begin{table}[h!]
\begin{center}
\begin{tabular}{|l|l|c|}
\hline
 {\bf Transferred from} & {\bf Transferred To} & {\bf Votes}\\
\hline
Begich & Peltola & 15,467 \\
\hline
Begich  & Palin & 27,053  \\
\hline
\end{tabular}
\end{center}
\caption{Transferred Votes}
\label{table:IRV-transfer}
\end{table}

\begin{table}[h!]
\begin{center}
\begin{tabular}{|l|c|c|}
\hline
 {\bf Candidate} & {\bf Votes} & {\bf Percentage}\\
\hline
Peltola & 91,266 & 51.48\% \\
\hline
Palin  & 86,026 & 48.52\%  \\
\hline
\end{tabular}
\end{center}
\caption{Second Round Special Election Results}
\label{table:IRV-round-2}
\end{table}

Instant Runoff Voting is often billed as a solution to many of the problems of traditional, ``choose one" plurality voting, also known as ``first past the post voting."  In plurality voting, voters may only vote for one candidate, and the candidate with the most votes wins, regardless of whether they win a majority of the votes.  With IRV, the winning candidate must receive a majority of the ballots that are still active in the final round.  Additionally, IRV can often eliminate the ``spoiler effect" seen in many plurality elections, where a candidate with only a small degree of support can attract votes that would otherwise have gone to a different candidate, thereby changing the outcome of the election.  With IRV, voters are frequently assured that they can safely rank their honest first choice candidate first, and then if that candidate is eliminated, they can still vote for their second choice candidate in the next round.  IRV is also widely claimed to reduce polarization by encouraging candidates to appeal to a wider variety of voters in order to attract second-place rankings from supporters of other candidates.

However, no voting system is perfect.  All of these supposed advantages of IRV can fail, and this election demonstrates how:

\begin{enumerate}
\item Peltola's final vote count of 91,266 represented 51.48\% of votes that were still active in Round 2, but this statistic fails to take into account that 11,290 ballots from Round 1 were exhausted and not counted in Round 2.   Peltola's 91,266 votes represent only 48.40\% of the 188,582 ballots that were active in Round 1. Thus the oft-repeated claim that IRV ``guarantees a majority for the winning candidate" is not true if by ``majority" we mean ``majority of all votes cast."  (It is important to note that this ``flaw" is not unique to IRV; {\em no} voting system can guarantee that the winner will receive a majority of all votes cast when there are more than two candidates.)
\item While voters who ranked Begich first had the opportunity to vote for their second-place candidate in Round 2, voters who ranked Palin or Peltola first never had their second-place vote considered.  In particular, voters who were assured that they could safely rank Palin first and still have their second-place vote for Begich counted if Palin were eliminated never got to express their support for Begich.
\item When the full Cast Vote Record was released, it became clear that a great deal of information about voter preferences was lost in the IRV tallying procedure.  Among voters expressing a second choice, Begich won an overwhelming majority of second-place votes---but these votes were never counted.  In fact, Begich was the {\em Condorcet winner} of this election: Based on the preferences expressed on the IRV ballots, Begich would have defeated both Palin and Peltola in head-to-head contests.  (More details will be given in Section \ref{details-sec}.)  This phenomenon, in which the Condorcet winner has broad support as a second choice candidate but is eliminated prior to the final round due to a lack of first-place votes, is referred to as a ``Condorcet failure."  When this happens, the candidates who remain in the final round are often the more polarizing candidates who have the strongest bases of first-choice supporters.

\end{enumerate}

This election displayed other anomalies of IRV as well, including the ``monotonicity paradox" and the ``no-show paradox;" see \cite{GSM22} for a more thorough discussion.

The goal of this paper is to explore how this election might have played out under two alternative voting systems: Approval Voting and STAR (``Score Then Automatic Runoff") Voting.  We will use the Cast Vote Record from the election to explore a range of possibilities for how the preferences expressed in these IRV ballots might translate to votes in each of these systems.

\section{Approval and STAR Voting}

Instant Runoff Voting is an {\em ordinal} (also known as a {\em ranked}) voting system: Voters rank candidates in order of preference.  Importantly, tied rankings are not allowed; if a voter assigns the same ranking to more than one candidate, that ballot is considered invalid and is not counted.  In this Alaska election, voters could express a tied ranking for their second- and third-place candidates by ranking only their first-place candidate and leaving the other two rankings blank, but voters had no way to express a tied ranking for their first- and second-place candidates.

\subsection{Approval Voting}

Approval Voting is the simplest example of a {\em cardinal} (also known as a {\em score} or {\em range}) voting system, in which voters give each candidate a score, the scores are added, and the candidate with the highest score wins the election.
In Approval Voting, the ballot is similar to a plurality voting ballot, except that voters may vote for as many candidates as they like, and the candidate who receives the most votes wins. (So voters effectively give each candidate a score of either 0 or 1.)  This can help eliminate the spoiler effect, as voters do not have to choose between voting for their true favorite candidate and a less favored candidate who is more likely to win.  

Approval Voting does require some strategic thinking on the part of voters: After voting for their favorite candidate and declining to vote for their least-favorite, is it better to vote for intermediate candidates in order to minimize the chance of a least-favorite candidate winning, or to decline to support them in order to maximize the chance of a favorite candidate winning?  If every voter decided to support only their first choice candidate---a strategy known as ``bullet voting"---then the outcome would be the same as in a plurality election.  In practice, the specific details of each election and voters' attitudes towards the various candidates may lead to a variety of outcomes.  

\subsection{STAR Voting}

STAR stands for ``Score Then Automatic Runoff;" it is a combination of cardinal and ordinal voting systems that was first introduced in 2014 \cite{star-creation}.  In STAR Voting, voters give each candidate a score, typically in the range of 0-5.  STAR ballots are tallied in two rounds: In the first round (the ``score" round), all scores are added and the candidates with the top two scores advance to the second round.  In the second round (the ``automatic runoff" round), every ballot that gives one of the two final candidates a higher score than the other counts as one vote in favor of that candidate, just as in a standard plurality election between two candidates.  (Any ballot that gave the final two candidates the same score is recorded as ``no preference" in the runoff.)  The winner of the runoff wins the election.

STAR Voting allows voters a greater range of expression than either IRV (unless there are more than 6 candidates) or Approval Voting, as they can more fully express their degree of support for each candidate.  Voters can express a range of ranked preferences by giving different scores to different candidates, and they can indicate tied preferences by giving the same score to multiple candidates.  The runoff round incentivizes voters to use the full range of scores available (unless they truly have no preference between candidates), so that their vote will be counted in the runoff round.

\subsection{Condorcet winners and losers}

Since IRV allows voters to rank all candidates, it is possible to determine from the Cast Vote Record how each voter would (presumably) vote in a theoretical head-to-head matchup between any pair of candidates, and consequently which candidate would win any such head-to-head election.  If there is a candidate who would defeat all other candidates in head-to-head elections, that candidate is called the {\em Condorcet winner} of the election. In practice, IRV elects the Condorcet winner most of the time, but as the Alaska election shows, this result is not guaranteed.  This was also famously the case in an IRV election for Mayor in Burlington, VT in 2009; see, e.g., \cite{ON14} for details.  

Similarly, if there is a candidate who would lose all head-to-head elections, that candidate is called the {\em Condorcet loser} of the election.  In this Alaska election, Begich was the Condorcet winner and Palin was the Condorcet loser.\footnote{Note that it was not possible to determine the Condorcet winner/loser from the official election reporting; this determination requires a full knowledge of the Cast Vote Record, including the record of second choice preferences for Palin and Peltola voters.}

While IRV does not guarantee the election of the Condorcet winner, it {\em does} guarantee that the Condorcet loser will {\em not} be elected. The Condorcet loser {\em may} survive until the final round---as indeed happened in the Alaska election---but by definition will lose to whichever other candidate survives until the final round.

By virtue of the automatic runoff step, STAR Voting also guarantees that the Condorcet loser will not be elected.  In Approval Voting, however, this result is theoretically {\em not} guaranteed.  While it is extremely unlikely that the Condorcet loser would win any particular Approval Voting election, we will see in Section \ref{approval-sec} how it is mathematically possible that, under precisely the right (or wrong?) circumstances, Palin could have won the Alaska special election if it had been conducted by Approval Voting.

\section{Cast Vote Record analysis for the IRV special election}\label{details-sec}

\begin{figure}[h!]
	\begin{center}
		\includegraphics[width=5in]{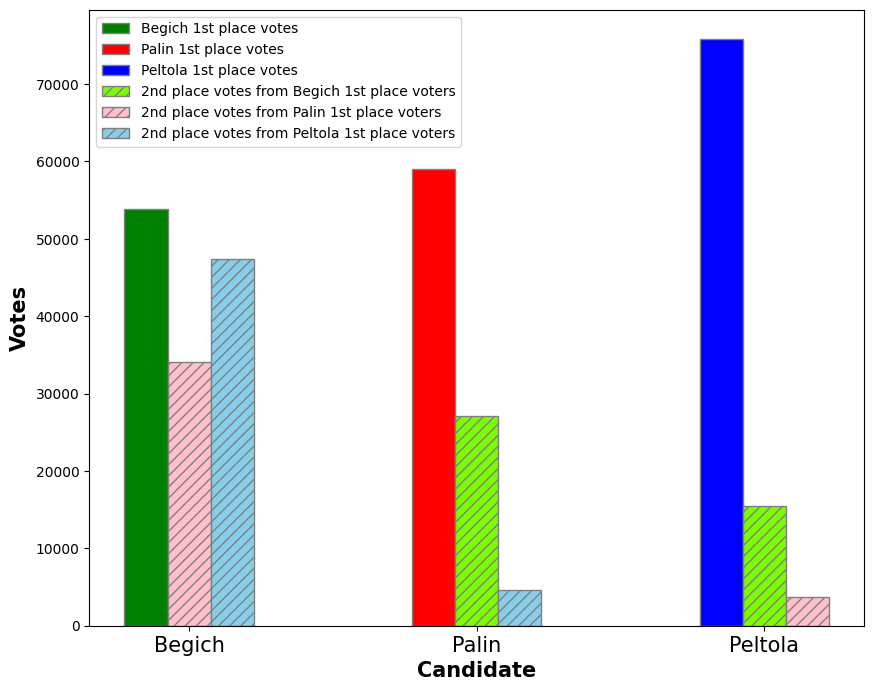}
	\end{center}
	\caption{First and Second Place Votes in the Special Election}
	\label{fig:CVR-ranked-votes-IRV}
\end{figure}

To analyze the Cast Vote Record,\footnote{Cast Vote Record obtained from https://www.elections.alaska.gov/election-results/e/?id=22sspg, accessed Nov. 4, 2022.}
we first applied the tabulation procedure described at 
\newline {\tt https://www.elections.alaska.gov/election-results/} under ``Sample RCV Report and Definitions" to the ballots in the Cast Vote Record to obtain a full profile of voters' preferences for the original IRV election.  The tabulation procedure described there leaves some ambiguity about how some ballots in the Cast Vote Record should be counted; in such cases we made choices that produced a tabulation that agrees with the published election results. The results of this tabulation, including all voters' preferences between the candidates, are shown in Table \ref{table:CVR-ranked-votes-IRV} and depicted graphically in Figure \ref{fig:CVR-ranked-votes-IRV}; we note that despite the ambiguity in the tabulation procedure, our numbers agree with those obtained independently by Graham-Squire and McCune in \cite{GSM22}.

\begin{table}[h!]
\begin{center}
\begin{tabular}{|l||c|c|c|c|c|c|c|c|c|}
\hline
1st place & Begich & Begich & Begich & Palin & Palin & Palin & Peltola & Peltola & Peltola \\
\hline
2nd place & --- & Palin & Peltola & --- & Begich & Peltola & --- & Begich & Palin \\
\hline
3rd place & --- & Peltola & Palin & --- & Peltola & Begich & --- & Palin & Begich \\
\hline\hline
Number of ballots & 11,290 & 27,053 & 15,467 & 21,272 & 34,049 & 3,652 & 23,747 & 47,407 & 4,645 \\
\hline
\end{tabular}
\end{center}
\caption{Voter Preferences in the Special Election}
\label{table:CVR-ranked-votes-IRV}
\end{table}

We can use the data in Table \ref{table:CVR-ranked-votes-IRV} to compute the results of theoretical head-to-head matchups for all three pairs of candidates:
\begin{itemize}
\item {\bf Begich vs. Palin:} Begich would receive votes from all ballots that ranked him first, plus votes that ranked Peltola first and Begich second.  Likewise, Palin would receive votes from all ballots that ranked her first, plus votes that ranked Peltola first and Palin second.  Thus we would have:
\begin{itemize}
\item Begich: 11,290 + 27,053 + 15,5467 + 47,407 = 101,217
\item Palin: 21,272 + 34,049 + 3,652 + 4,645 = 63,618
\end{itemize}
So Begich would defeat Palin with approximately 61.41\% of the vote.

\item {\bf Begich vs. Peltola:} Begich would receive votes from all ballots that ranked him first, plus votes that ranked Palin first and Begich second.  Likewise, Peltola would receive votes from all ballots that ranked her first, plus votes that ranked Palin first and Peltola second.  Thus we would have:
\begin{itemize}
\item Begich: 11,290 + 27,053 + 15,5467 + 34,049 = 87,859
\item Peltola: 23,747 + 47,407 + 4,645 + 3,652 = 79,451 
\end{itemize}
So Begich would defeat Peltola with approximately 52.51\% of the vote.

\item {\bf Palin vs. Peltola:} Palin would receive votes from all ballots that ranked her first, plus votes that ranked Begich first and Palin second.  Likewise, Peltola would receive votes from all ballots that ranked her first, plus votes that ranked Begich first and Peltola second.  Thus we would have:
\begin{itemize}
\item Palin: 21,272 + 34,049 + 3,652 + 27,053 = 86,026
\item Peltola: 23,747 + 47,407 + 4,645 + 15,467 = 91,266
\end{itemize}
So Peltola would defeat Palin with approximately 51.48\% of the vote. (And in fact, this is exactly what happened in the final round after Begich was eliminated.)

\end{itemize}
These calculations show that for this election, Begich was the Condorcet winner and Palin was the Condorcet loser.

In the next two sections, we will explore a range of possibilities for how this election might have played out under either Approval or STAR Voting.

\section{Possible election outcomes with Approval Voting}\label{approval-sec}

In order to model an Approval Voting election based on the Cast Vote Record, we make the following assumptions:
\begin{itemize}
\item The 56,309 voters who ranked only one candidate will vote for that candidate and will not vote for the remaining two candidates.
\item The 132,273 voters who ranked all three candidates will vote for their first-place candidate and will not vote for their third-place candidate. We will consider a range of possibilities for what percentage of these voters choose to vote for their second-place candidate.
\end{itemize}

\begin{figure}[h!]
	\begin{center}
		\includegraphics[width=6.4in]{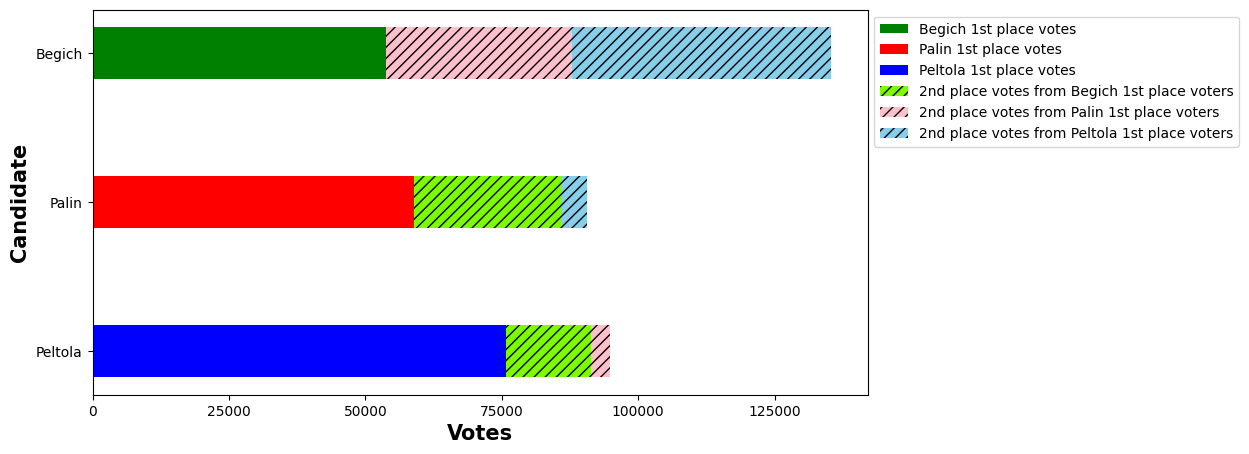}
	\end{center}
	\caption{Range of Possible Approval Voting Outcomes for the Special Election}
	\label{fig:approval-voting-outcomes}
\end{figure}

The range of possible outcomes for this election under this model is depicted numerically in Table \ref{table:approval-voting-outcomes} and graphically in
Figure \ref{fig:approval-voting-outcomes}.  For each candidate, the dark-colored bar in Figure \ref{fig:approval-voting-outcomes} indicates their minimum level of support, based on first-place votes.  The light-colored, shaded bars indicate the range of support that is potentially available from second-place votes, color-coded by the range available from each of the other candidates' first-place voters.

\begin{table}[h!]
\begin{center}
\begin{tabular}{|l||c|c|}
\hline
& {\bf Minimum votes} & {\bf Maximum votes} \\
\hline
Begich & 53,810 & 135,266  \\
\hline
Palin & 58,973 & 90,671 \\
\hline
Peltola & 75,799 &  94,918   \\
\hline
\end{tabular}
\end{center}
\caption{Range of Possible Approval Voting Outcomes for the Special Election}
\label{table:approval-voting-outcomes}
\end{table}

This chart shows that, in theory, any of the three candidates could win the election under exactly the right circumstances.
\begin{itemize}
\item {\bf How Begich could win:} If a substantial portion of the Peltola and/or Palin voters who ranked Begich second opted to vote for him as well, then Begich could win easily, regardless of how many votes Palin and Peltola received from second-place rankings.  Indeed, even if Begich received {\em no} votes from voters who ranked him second to Palin, he would need votes from only 41,109 of the 47,407 voters who ranked him second to Peltola (about 87\%) in order to exceed the {\em maximum} possible number of votes for either Palin or Peltola.

\item {\bf How Peltola could win:} If all voters opted to bullet vote, then the outcome would be the same as for a plurality election and Peltola would win.  Alternatively, if most Peltola and Palin voters opted to bullet vote, while most Begich voters opted to vote for their second-choice candidate, the outcome would likely be similar to the final total in the IRV election, and Peltola would likely still win the election.

\item {\bf How Palin could win:} If all voters declined to support a second-choice candidate from the opposite party of their first-choice candidate, then Peltola would receive no support from any second-place rankings.  In this case, Begich and Palin would receive no votes from Peltola's first-place voters, but they could still receive support from each other's first-place voters.  If most voters who ranked Begich second to Palin declined to support Begich while most voters who ranked Palin second to Begich opted to support Palin, it is theoretically possible that Palin could win the election.

\end{itemize}

Now, how likely are each of these scenarios?  The outcomes hinge on what percentage of voters with each preference profile choose to vote for their second-place candidate, and this is difficult to predict in advance.
Approval Voting has a limited track record in real-world political elections, with the best-known Approval Voting elections taking place in nonpartisan primary elections for municipal offices in St. Louis, MO and nonpartisan municipal elections in Fargo, ND since 2020.  Results from these elections are available at {\tt https://approval.vote/}; here is a summary:

\begin{itemize}
\item In Fargo, Approval Voting was used in 2020 and 2022 to elect two City Commissioners in a single election each year, and in 2022 to elect the Mayor. For the City Commissioner elections, voters approved an average of 2.3 candidates (out of 7 running) in 2020 and an average of 3.1 candidates (out of 15 running) in 2022.  For the 2022 Mayoral election, voters approved an average of 1.6 candidates (out of 7 running).

\item In St. Louis, Approval Voting was used in 2021 for two-winner primary elections for Mayor, Comptroller, and 16 ward-based Aldermen.  Of these, only the Mayoral and 7 of the Alderman elections featured 3 or more candidates.  In each of the 6 Alderman elections with exactly 3 candidates, voters approved averages of 1.1 or 1.2 candidates; in the Alderman election with 6 candidates, voters approved an average of 1.4 candidates, and in the Mayoral election with 4 candidates, voters approved an average of 1.6 candidates.  Note that in all of these elections, the average number of approvals was {\em less} than the number of winning candidates (i.e., two).
\end{itemize}

While it is certainly not a given that a statewide, partisan election would follow the same pattern as a local, nonpartisan election, it appears that on average, voters generally prefer to support a relatively small number of candidates in an Approval Voting election.  

For our hypothetical election, if we assume an extremely simple (and admittedly unrealistic!) model where the percentage of voters who choose to vote for their second-choice candidate is the same for all candidate ranking patterns, then the threshold required for Begich to overtake Peltola is about 1.36 approvals per ballot.  So in this model, if at least 36\% of voters chose to vote for their second-ranked candidate, then Begich would win the election; otherwise Peltola would win.  (Palin would never win with this model, as the scenarios under which she could win would require her to receive many more second-place votes than the other two candidates.)  Based on the data from Fargo and St. Louis, it seems unlikely---although certainly not impossible---that this threshold would be reached, and therefore likely that Peltola would still win the election.  But there are certainly realistic scenarios under which either Begich or Peltola might win.

\section{Possible election outcomes with STAR Voting}\label{star-sec}

In order to model a STAR Voting election based on the Cast Vote Record, we make the following assumptions.  We start by assuming that voters will maximize the impact of their votes by giving their favorite candidate(s) 5 stars and their least favorite candidate(s) 0 stars.
\begin{itemize}
\item The 56,309 voters who ranked only one candidate will give that candidate 5 stars and will give the remaining two candidates 0 stars.
\item The 132,273 voters who ranked all three candidates will give their first-place candidate 5 stars and will give their third-place candidate 0 stars. We will consider a range of possibilities for how may stars, on average, these voters choose to give their second-place candidate.  
\end{itemize}
There is one key difference between our models for STAR and Approval Voting: Since voters who ranked all three candidates indicated a preference between their second- and third-place candidates, for STAR Voting we will assume that they will give their second-place candidate {\em at least} one star, so as to maintain their preference order for the runoff round.  And, while there is the possibility that some of these voters might choose to give 5 stars to each of their top two candidates, the runoff round is a sufficiently strong disincentive to do so that we will assume that they will give their second-place candidate {\em no more than}
4 stars.

\begin{figure}[h!]
	\begin{center}
		\includegraphics[width=6.4in]{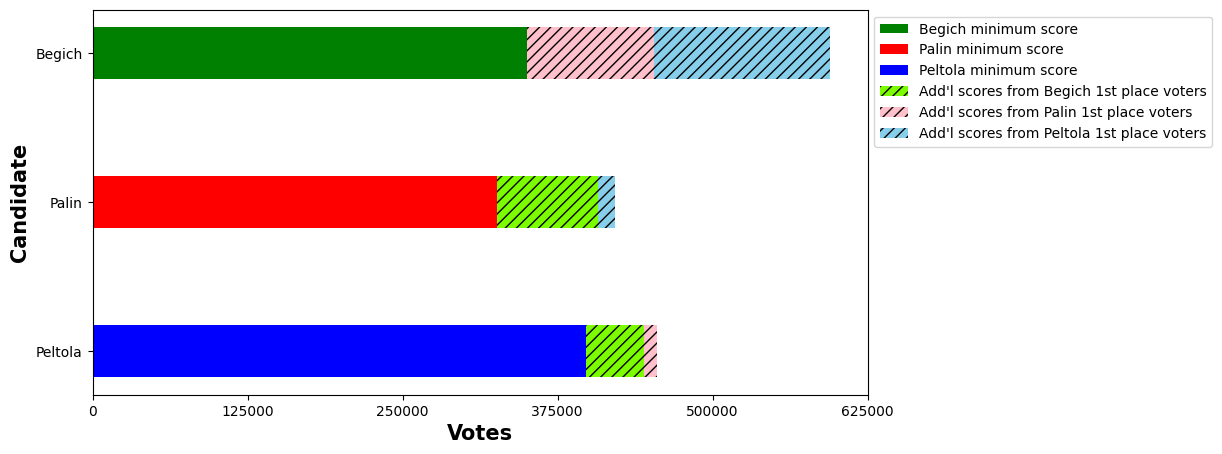}
	\end{center}
	\caption{Range of Possible STAR Voting Outcomes for the Special Election}
	\label{fig:STAR-voting-outcomes}
\end{figure}

The range of possible outcomes for the score round of this election under this model is depicted numerically in Table \ref{table:STAR-voting-outcomes} and graphically in
Figure \ref{fig:STAR-voting-outcomes}.
For each candidate, the dark-colored bar in Figure \ref{fig:STAR-voting-outcomes} indicates their minimum level of support, based on 5 stars from each first-place vote and 1 star from each second-place vote.  The light-colored, shaded bars indicate the range of support that is potentially available from additional stars for second-place votes, color-coded by the range available from each of the other candidates' first-place voters.

\begin{table}[h!]
\begin{center}
\begin{tabular}{|l||c|c|}
\hline
& {\bf Minimum score} & {\bf Maximum score} \\
\hline
Begich & 350,506 &  594,874 \\
\hline
Palin & 326,563 &  421,657 \\
\hline
Peltola & 398,114 & 455,471  \\
\hline
\end{tabular}
\end{center}
\caption{Range of Possible STAR Voting Outcomes for the Special Election}
\label{table:STAR-voting-outcomes}
\end{table}

Comparing with Figure \ref{fig:approval-voting-outcomes} shows some important differences between the scoring outcomes for STAR and Approval Voting.  In Approval Voting, second-place rankings can contribute anywhere from 0\%-100\% as much as first-place rankings.  But in our STAR Voting model, the assumption that second-place rankings receive between 1 and 4 stars means that this range is reduced to 20\%-80\%.  This 20\% floor pays off handsomely for Begich, who received many more second-place rankings than either of the other two candidates.
\begin{itemize}
\item {\bf How Begich could win:} Since Begich is the Condorcet winner, he would win the election as long as he made it to the runoff step. If the Peltola and Palin voters who ranked Begich second (81,456 voters total) gave him an average of at least 1.88 stars instead of the assumed minimum of 1 star, that would give him at least 71,681 additional stars, for a total of at least 422,187 stars---just above Palin's theoretical {\em maximum} of 421,657 stars.  Absent a situation in which Begich's second-place voters gave him many fewer stars on average than Palin's second-place voters gave her, Begich would advance to the runoff and win the election regardless of how many stars Palin and Peltola received from second-place voters.

\item {\bf How Peltola could win:} Even if all voters gave their second-choice candidate the minimum of 1 star, Peltola would receive the highest score in the first round but would still lose the runoff to Begich.  Peltola's only path to win the election would be for most of Begich's second-place voters to give him only 1 star and for a significant fraction of Palin's second-place voters to give her 3 or 4 stars, thereby allowing Palin to defeat Begich in the score round and advance to the runoff, where Peltola would defeat Palin.

\end{itemize}

As the Condorcet loser, Palin could not win this election under any circumstances.

Now, how likely are each of these scenarios?  
STAR Voting has an even more limited track record than Approval Voting in real-world political elections.  As of this writing, it has only been used for some internal party elections in Oregon, although an effort is underway to introduce a 2024 ballot initiative to adopt STAR Voting statewide in Oregon.  (More information is available from the Equal Vote Coalition at {\tt https://www.equal.vote/}.)

But even with the uncertainty about how voters might choose to vote in a STAR Voting election, it seems clear that Begich has a much stronger path to victory here than in an Approval Voting election.  The only scenario in which Begich might lose would require a substantial asymmetry in the way that voters with different preferences choose to score their second-place candidates, with many more Begich voters choosing to score Palin highly than the reverse, and essentially no Peltola voters choosing to score Begich highly.  While this is theoretically possible, it seems extremely unlikely; thus we conclude that Begich would almost certainly win the election.

\section{Postscript: A do-over?}

In a highly unusual turn of events, the same three candidates from the special election (plus one additional candidate, Libertarian Chris Bye, who was eliminated in the first round) faced off again just a few months later in the November 2022 regular general election for U.S. House.  Some observers wondered whether a significant number of Republican voters might take a lesson from the special election and switch their first-place votes from Palin to Begich in order to avoid a repeat victory by Peltola; however, this emphatically did {\em not} happen. The regular election played out similarly to the special election, with Begich being eliminated before Palin despite strong second-place support, and Palin losing to Peltola in the final round.  Round-by-round results as reported by the Alaska Division of Elections are shown in Table \ref{table:general-election-official-results}.\footnote{Official results and Cast Vote Record for the general election obtained from https://www.elections.alaska.gov/election-results/e/?id=22genr, accessed March 20, 2024.}

\begin{table}[h!]
\begin{center}
\begin{tabular}{|l|c|c|c|}
\hline
 {\bf Candidate} & {\bf Round 1} & {\bf Round 2} & {\bf Round 3} \\
\hline
Peltola &  128,755 (48.66\%)  &  129,786 (49.22\%) & 137,263 (54.96\%) \\
\hline
Palin  &  68,330 (25.82\%) &  69,399 (26,32\%)  & 112,471 (45.04\%) \\
\hline
Begich  & 62,505 (23.62\%) &  64,499 (24.46\%)  & --- \\
\hline
Bye & 4,999 (1.89\%) &  --- &  --- \\
\hline
\end{tabular}
\end{center}
\caption{General Election Results}
\label{table:general-election-official-results}
\end{table}

Although the order of candidate elimination and the final result were the same as for the special election, there were significant differences between the special and general elections.  Most importantly, Peltola's support was much stronger in the general election than in the special election: she won 48.66\% of the first-round votes, compared with 40.19\% in the special election.  There was also a sizable shift in the distribution of transfers of Begich's votes to Palin and Peltola: in the special election, these transfers split about 64\%-36\% for Palin, but in the special election, these transfers split about 85\%-15\% for Palin.

\begin{figure}[h!]
	\begin{center}
		\includegraphics[width=5in]{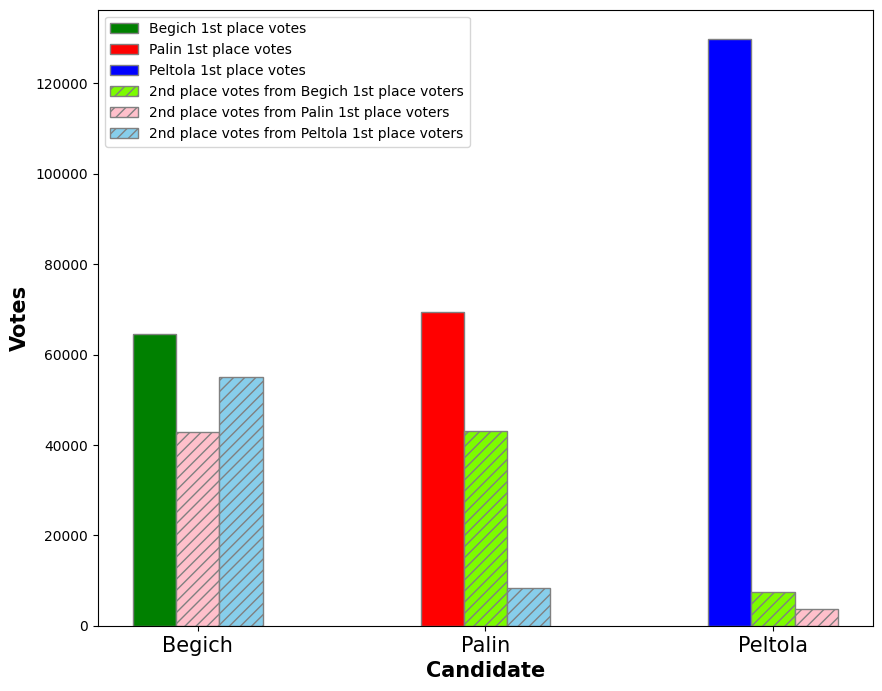}
	\end{center}
	\caption{First and Second Place Votes in the General Election}
	\label{fig:CVR-ranked-votes-IRV-general}
\end{figure}

As in the special election, an analysis of the Cast Vote Record is required in order to more fully explore voters' preferences among the candidates.
Voters' preferences among the three main candidates as recorded in the Cast Vote Record (ignoring preferences for Bye) are shown in Table \ref{table:CVR-ranked-votes-IRV-general} and depicted graphically in Figure \ref{fig:CVR-ranked-votes-IRV-general}.  (Compare with the results from the special election shown in Table \ref{table:CVR-ranked-votes-IRV} and Figure \ref{fig:CVR-ranked-votes-IRV}.)  These results also reveal some differences between the special and general elections that are not evident from the official election reporting:

\begin{itemize}
\item Most importantly, Peltola was the Condorcet winner of the general election, so this election did not exhibit the Condorcet failure seen in the special election. Head-to-head matchups computed from the data in Table \ref{table:CVR-ranked-votes-IRV-general} show that Peltola would defeat Palin with approximately 54.96\% of the vote, and Peltola would defeat Begich with approximately 55.41\% of the vote.
\item There was less crossover voting between political parties in the general election than in the special election.  On the Democratic side, about 51.12\% of Peltola's second-round voters (after Bye had been eliminated) declined to rank either Begich or Palin in the general election, compared to 31.32\% in the special election.  In particular, the percentage of Peltola voters who chose to rank Begich over Palin declined from 62.38\% in the special election to 42.39\% in the general election.  Meanwhile, on the Republican side the percentage of Begich voters who chose to rank Peltola over Palin declined from 28.74\% in the special election to 11.59\% in the general election, while the percentage of Begich voters who chose to rank Palin over Peltola increased from 50.28\% in the special election to 66.78\% in the general election.  Second-place support for Peltola among Palin voters also decreased slightly, from 6.19\% in the special election to 5.34\% in the general election.

\end{itemize}

\begin{table}[h!]
\begin{center}
\begin{tabular}{|l||c|c|c|c|c|c|c|c|c|}
\hline
1st place & Begich & Begich & Begich & Palin & Palin & Palin & Peltola & Peltola & Peltola \\
\hline
2nd place & --- & Palin & Peltola & --- & Begich & Peltola & --- & Begich & Palin \\
\hline
3rd place & --- & Peltola & Palin & --- & Peltola & Begich & --- & Palin & Begich \\
\hline\hline
Number of ballots & 13,950 & 43,072 & 7,477 & 22,751 & 42,941 & 3,707 & 66,349 & 55,013 & 8,424 \\
\hline
\end{tabular}
\end{center}
\caption{Voter Preferences in the General Election}
\label{table:CVR-ranked-votes-IRV-general}
\end{table}

Not surprisingly, Peltola also performs much better in the general election under our models for Approval and STAR voting.  Applying the same analysis as in Sections  \ref{approval-sec} and \ref{star-sec} to the numbers in Table \ref{table:CVR-ranked-votes-IRV-general} shows that:
\begin{itemize}
\item For Approval Voting, Palin cannot win under any circumstances because the maximum number of approvals that she could receive (120,895) is below Peltola's minimum number of approvals (129,786).  Begich could win, but he would have a higher hill to climb than in the special election.  If we use the simple model where we assume that the percentage of voters who choose to vote for their second-choice candidate is the same for all candidate ranking patterns, then the threshold required for Begich to overtake Peltola is about 1.76 approvals per ballot (compared to 1.36 for the special election).  Even if Peltola received no votes at all from Begich's or Palin's voters, Begich would still need about 67\% of all his possible votes from Palin's and Peltola's voters in order to defeat Peltola. (In particular, he would need about 40\% of all his possible votes from Peltola voters, even if he received 100\% of all his possible votes from Palin voters.)  While theoretically possible, this seems very unlikely, so we conclude that Peltola would probably have won the election under Approval Voting.
\item For STAR voting, Peltola is guaranteed to win: Peltola's minimum number of stars (660,114) is larger than Palin's maximum (552,979), so Peltola is guaranteed to make it to the runoff round, which she would then win by virtue of being the Condorcet winner.

\end{itemize}

\section{Conclusions}

When implementing a new and unfamiliar voting system, it is almost impossible to predict in advance how voters might navigate the new procedures.  Extensive voter education should be an essential component of implementation, but this is challenging and takes time.  Voter behavior is likely to evolve over time as voters become more familiar with a new system and gain understanding as to how to use it to express their preferences most effectively.  It will be interesting to see how Approval Voting patterns evolve over time in St. Louis and Fargo as voters become more comfortable with it---and perhaps even more interesting to see how voters respond to STAR Voting if it is adopted in Oregon in 2024.

That said, with the limited information that we have regarding voter behavior in Approval and STAR Voting elections, it seems somewhat likely that for the special election, Peltola would still have won under Approval Voting, while Begich---the Condorcet winner---would most likely have won the election under STAR Voting.  For the general election, Peltola would almost certainly have won regardless of which system was used.

We close by emphasizing that, while the special election demonstrates that anomalies such as the Condorcet failure can and do happen in IRV elections, they are very rare in practice.  The general election displays a much more typical pattern, in which the winner of the election is also the Condorcet winner.  In \cite{GSM23}, Graham-Squire and McCune analyzed a database of 182 American elections conducted by IRV between 2004 and 2022, and they found only two elections (the Alaska special election and the 2009 mayoral election in Burlington, VT) that failed to elect the Condorcet winner---plus one additional election for which there was no Condorcet winner.  

Nevertheless, these rare failures can have consequences. Following the 2009 Burlington mayoral election, the city of Burlington repealed the use of Ranked Choice Voting and returned to plurality elections.  Advocates for Ranked Choice Voting and other alternative voting systems need to be realistic in their promises to voters about what these systems can accomplish, and also about their potential flaws, however rare they may be.

\section{Acknowledgments}
The author would like to thank Neal McBurnett, David McCune, and Daryl Deford for their assistance with processing the Cast Vote Records.  This work would not have been possible without their assistance, and I am very grateful for their generous sharing of their expertise.

\bibliographystyle{amsplain}
\bibliography{AK_Special_2022_bib}

\end{document}